\documentclass[12pt]{article}
\usepackage{epsfig}
\usepackage{subfigure}
\usepackage{amsmath}
\usepackage{amsfonts}
\usepackage{amssymb}
\usepackage{url}
\usepackage{hyperref}
\newcommand{\be}{\begin{equation}}
\newcommand{\ee}{\end{equation}}
\newcommand{\bea}{\begin{eqnarray}}
\newcommand{\eea}{\end{eqnarray}}

\begin{document}
\begin{titlepage}
\begin{flushright}
KEK-TH-1137\\
\end{flushright}
\vspace{4\baselineskip}
\begin{center}
{\Large\bf 
Solving problems of 4D minimal SO(10) model \\
    in a warped extra dimension             
}
\end{center}
\vspace{1cm}
\begin{center}
{\large Takeshi Fukuyama $^{a,}$
\footnote{\tt E-mail:fukuyama@se.ritsumei.ac.jp},
Tatsuru Kikuchi $^{b,}$
\footnote{\tt E-mail:tatsuru@post.kek.jp}
and Nobuchika Okada $^{b,c,}$
\footnote{\tt E-mail:okadan@post.kek.jp}}
\end{center}
\vspace{0.2cm}
\begin{center}
${}^{a}$ {\small \it Department of Physics, Ritsumeikan University,
Kusatsu, Shiga, 525-8577, Japan}\\[.2cm]
${}^{b}$ {\small \it Theory Division, KEK,
Oho 1-1, Tsukuba, Ibaraki, 305-0801, Japan}\\
${}^{c}$ {\small \it Department of Particle and Nuclear Physics,
The Graduate University for Advanced Studies, \\
Oho 1-1, Tsukuba, Ibaraki, 305-0801, Japan}\\
\medskip
\vskip 10mm
\end{center}
\vskip 10mm
\begin{abstract}
The minimal renormalizable supersymmetric SO(10) model, 
 an SO(10) framework with only one ${\bf 10}$ and 
 one $\overline{\bf 126}$ Higgs multiplets 
 in the Yukawa sector, 
 is attractive because of its high predictive power 
 for the neutrino oscillation data.  
However, this model suffers from problems 
 related to running of gauge couplings. 
The gauge coupling unification may be spoiled 
 due to the presence of Higgs multiplets 
 much lighter than the grand unification (GUT) scale. 
In addition, the gauge couplings blow up around the GUT scale 
 because of the presence of Higgs multiplets 
 of large representations. 
We consider the minimal SO(10) model in the warped extra dimension 
 and find a possibility to solve these problems. 
\end{abstract}
\end{titlepage}

\section{Introduction}

A particularly attractive idea for the physics beyond 
 the standard model (SM) is the possibility 
 of grand unified theory (GUT). 
In the context of GUTs 
 the diverse set of particle representations 
 and parameters in the SM are unified into 
 a simple and more predictive framework. 
 From this unified picture, one can explain, for example, 
 quantization of electric charges of quarks and leptons. 
Current experimental data for the standard model 
 gauge coupling constants suggest 
 the successful gauge coupling unification 
 in the minimal supersymmetric standard model (MSSM), 
 and thus strongly support 
 the emergence of a supersymmetric (SUSY) GUT 
 around $M_{\rm GUT} \simeq 2 \times 10^{16}$ GeV.

Among several GUTs, a model based on the gauge group SO(10) 
 is particularly attractive in the following reason. 
SO(10) is the smallest simple gauge group 
 under which the entire SM matter content of each generation 
 is unified into a single anomaly-free irreducible representation, 
 namely the spinor ${\bf 16}$ representation. 
In fact, this ${\bf 16}$ representation includes 
 right-handed neutrino and 
 SO(10) GUT incorporates the see-saw mechanism \cite{see-saw}  
 that can naturally explain the lightness of the light neutrino masses.

Among several models based on the gauge group SO(10), 
 the so-called minimal SO(10) model has been paid 
 a particular attention, where two Higgs multiplets 
 $\{{\bf 10} \oplus {\bf \overline{126}}\}$ 
 are utilized for the Yukawa couplings with matters 
 ${\bf 16}_i~(i=\mbox{generation})$ \cite{Babu:1992ia}. 
A remarkable feature of the model is its high predictive power 
 of the neutrino oscillation parameters 
 as well as reproducing charged fermion masses and mixing angles. 
It has been pointed out that 
 CP-phases in the Yukawa sector play an important role 
 to reproduce the neutrino oscillation data \cite{Matsuda:2000zp}. 
More detailed analysis incorporating the renormalization group (RG) 
 effects in the context of MSSM \cite{Fukuyama:2002ch} 
 has explicitly shown that the model is consistent with the neutrino
 oscillation data.

However, after KamLAND data \cite{Eguchi:2002dm} was released, 
 the results in Ref.~\cite{Fukuyama:2002ch} were found to be deviated 
 by 3$\sigma$ from the observations. 
Afterward this minimal SO(10) was modified by many authors, 
 using the so-called type-II see-saw mechanism \cite{Bajc:2002iw} 
 and/or considering a ${\bf 120}$ Higgs coupling to the matter
 in addition to the ${\bf \overline{126}}$ Higgs \cite{Goh:2003hf}. 
Based on an elaborate input data scan \cite{Babu:2005ia, Bertolini:2006pe}
 it has been shown that the minimal SO(10) is essentially consistent 
 with low energy data of fermion masses and mixing angles.

On the other hand, it has been long expected to construct 
 a concrete Higgs sector of the minimal SO(10) model. 
A simplest and renormalizable Higgs superpotential 
 was constructed explicitly and the patterns of 
 the SO(10) gauge symmetry breaking to the standard model one  
 was shown \cite{Parida, Fukuyama:2004xs, Bajc:2004xe, Aulakh:2004hm}.
This construction gives some constraints among the vacuum expectation
 values (VEVs) of several Higgs multiplets, 
 which give rise to a trouble in the gauge coupling unification. 
The trouble comes from the fact that the observed neutrino oscillation
 data suggests the right-handed neutrino mass around $10^{12-13}$ GeV,
 which is far below the GUT scale. 
This intermediate scale is provided by Higgs field VEV, 
 and several Higgs multiplets are expected to have their masses 
 around the intermediate scale and contribute to 
 the running of the gauge couplings. 
Therefore, the gauge coupling unification 
 at the GUT scale may be spoiled. 
This fact has been explicitly shown in Ref.~\cite{Bertolini:2006pe}, 
 where the gauge couplings are not unified any more 
 and even the ${\rm SU}(2)$ gauge coupling blows up below the GUT scale. 
In order to avoid this trouble and keep the successful gauge coupling 
 unification as usual, it is desirable that all Higgs multiplets 
 have masses around the GUT scale, but some Higgs fields 
 develop VEVs at the intermediate scale. 
More Higgs multiplets and some parameter tuning in the Higgs sector 
 are necessary to realize such a situation. 

In addition to the issue of the gauge coupling unification, 
 the minimal SO(10) model potentially suffers from the problem 
 that the gauge coupling blows up around the GUT scale. 
This is because the model includes many Higgs multiplets of 
 higher dimensional representations. 
In field theoretical point of view, this fact implies 
 that the GUT scale is a cutoff scale of the model, 
 and more fundamental description of the minimal SO(10) model 
 would exist above the GUT scale.

In order to solve these problems related 
 to the gauge coupling running, 
 we can consider two possibilities. 
One is to replace Higgs fields of large representations 
 into smaller ones and to provide Yukawa couplings 
 in the original minimal SO(10) model 
 as higher dimensional operators \cite{Chang:2004pb}.  
In this way, we can keep the gauge couplings 
 in perturbative regime until the Planck scale or the string scale. 
The other possibility is what we explore in this letter: 
 to provide the GUT scale as the cutoff scale 
 of effective filed theory in a natural way.  
For our purpose, we embed the minimal SO(10) model 
 into a warped extra dimension model \cite{RS}. 
In this scenario, the warped metric give rise to 
 an effective cutoff in 4-dimensional effective theory, 
 which is warped down to a low scale from the fundamental mass scale 
 of the original model (a higher dimensional Planck scale). 
We choose appropriate model parameters so as to realize 
 the effective cutoff scale as the GUT scale. 
Furthermore, in the contest of a warped extra dimension 
 we can propose a simple setup that naturally generates 
 right-handed neutrino masses at intermediate scale 
 even with Higgs field VEVs at the GUT scale. 
Thus, the gauge coupling unification remains 
 as usual in the MSSM.

This paper is organized as follows: 
In the next section, we give a brief review of 
 the minimal SUSY SO(10) model, and claim the problems 
 related to the running of the gauge couplings. 
In section 3, we construct a minimal SO(10) model 
 in the contest of the warped extra dimension 
 and propose a simple setup that can solves the problems. 
The last section is devoted to summary.

\section{Minimal supersymmetric SO(10) model} 
We begin by giving a brief review on the minimal SUSY SO(10) model 
 and show the GUT relation among fermion mass matrices. 
Even when we concentrate our discussion on the issue how to 
 reproduce the realistic fermion mass matrices in the SO(10) model, 
 there are lots of possibilities for introduction of Higgs multiplets.  
The minimal supersymmetric SO(10) model is the one where only 
 one {\bf 10} and one $\overline{\bf 126}$ Higgs multiplets 
 have Yukawa couplings with ${\bf 16}$ matter multiplets such as 
\begin{eqnarray}
W = Y_{10}^{ij} {\bf 16}_i {\bf 10}_H {\bf 16}_j 
+ Y_{126}^{ij} {\bf 16}_i {\bf \overline{126}}_H {\bf 16}_j \; , 
\label{Yukawa1}
\end{eqnarray} 
where ${\bf 16}_i$ is the matter multiplet of the $i$-th generation, 
 ${\bf 10}_H$ and ${\bf \overline{126}}_H$ are the Higgs multiplet of 
 ${\bf 10}$ and $\overline{\bf 126}$ representations under SO(10), 
 respectively.  
Note that, by virtue of the gauge symmetry, 
 the Yukawa couplings, $Y_{10}$ and $Y_{126}$, are complex symmetric 
 $3 \times 3$ matrices.  
We assume some appropriate Higgs multiplets, whose vacuum expectation 
 values (VEVs) correctly break the SO(10) GUT gauge symmetry 
 into the standard model one at the GUT scale, 
 $M_{\rm GUT} = 2 \times 10^{16}$ GeV.  
Suppose the Pati-Salam subgroup, 
 $G_{422}={\rm SU}(4)_c \times {\rm SU}(2)_L \times {\rm SU}(2)_R$, 
 at the intermediate breaking stage.  
Under this symmetry, the above Higgs multiplets are decomposed as 
 ${\bf 10} \rightarrow 
 ({\bf 6},{\bf 1},{\bf 1}) + ({\bf 1},{\bf 2},{\bf 2}) $ 
 and 
 $\overline{\bf 126} \rightarrow 
 ({\bf 6}, {\bf 1}, {\bf 1} ) 
 + ( {\bf \overline{10}}, {\bf 3}, {\bf 1}) 
 + ({\bf 10}, {\bf 1}, {\bf 3})  
 + ({\bf 15}, {\bf 2}, {\bf 2}) $, 
 while ${\bf 16} \rightarrow ({\bf 4}, {\bf 2}, {\bf 1}) 
 + (\overline{\bf 4}, {\bf 1}, {\bf 2})$.  
Breaking down to the standard model gauge group, 
 ${\rm SU}(4)_c \times {\rm SU}(2)_R  \rightarrow {\rm SU}(3)_c \times {\rm U}(1)_Y$,   
 is accomplished by non-zero VEV  of 
 the $({\bf 10}, {\bf 1}, {\bf 3})$ Higgs multiplet. 
Note that Majorana masses for the right-handed neutrinos 
 are also generated by this VEV through the Yukawa coupling 
 $Y_{126}$ in Eq.~(\ref{Yukawa1}).  
In general, the ${\rm SU}(2)_L$ triplet Higgs in 
 $({\bf \overline{10}}, {\bf 3}, {\bf 1}) \subset \overline{\bf 126}$ 
 would obtain the VEV induced through the electroweak symmetry breaking 
 and may play a crucial role 
 of the light Majorana neutrino mass matrix. 
This model is called the type-II see-saw model, 
 and we include this possibility in the following.

After the symmetry breaking, we find two pair of Higgs doublets 
 in the same representation as the pair in the MSSM.  
One pair comes from $({\bf 1},{\bf 2},{\bf 2}) \subset {\bf 10}$ 
 and the other comes from 
 $({\bf 15}, {\bf 2}, {\bf 2}) \subset \overline{\bf 126}$.  
Using these two pairs of the Higgs doublets, 
 the Yukawa couplings of Eq.~(\ref{Yukawa1}) are rewritten as 
\begin{eqnarray}
W_Y &=& (U^c)_i  \left(
Y_{10}^{ij}  H^u_{10} + Y_{126}^{ij}  H^u_{126} \right) Q
+ (D^c)_i  \left(
Y_{10}^{ij}  H^d_{10} + Y_{126}^{ij}  H^d_{126} \right) Q_j  
\nonumber \\ 
&+& (N^c)_i \left( 
Y_{10}^{ij}  H^u_{10} - 3 Y_{126}^{ij} H^u_{126} \right) L_j 
+ (E^c)_i  \left(
Y_{10}^{ij}  H^d_{10}  - 3 Y_{126}^{ij} H^d_{126} \right) L_j   
\nonumber \\
&+&
 L_i \left( Y_{126}^{ij} \; v_T \right) L_j +
(N^c)_i \left( Y_{126}^{ij} \; v_R \right) (N^c)_j \;  , 
\label{Yukawa2}
\end{eqnarray} 
where $U^c$, $D^c$, $N^c$ and $E^c$ are the right-handed ${\rm SU}(2)_L$ 
 singlet quark and lepton superfields, $Q$ and $L$ 
 are the left-handed ${\rm SU}(2)_L$ doublet quark and lepton superfields, 
 $H_{10}^{u,d}$ and $H_{126}^{u,d}$ are up-type and down-type 
 Higgs doublet superfields originated 
 from ${\bf 10}$ and ${\bf \overline{126}}$, respectively, 
 and the last two terms are the Majorana mass term of 
 the left-handed and the right-handed neutrinos, respectively, 
 developed by the VEV of 
 the $({\bf \overline{10}}, {\bf 3}, {\bf 1})$ Higgs ($v_T$) 
 and the $({\bf 10}, {\bf 1}, {\bf 3})$ Higgs ($v_R$).  
The factor $-3$ in the lepton sector is the Clebsch-Gordan (CG) 
 coefficient.

In order to keep the successful gauge coupling unification, 
 suppose that one pair of Higgs doublets 
 given by a linear combination $H_{10}^{u,d}$ and $H_{126}^{u,d}$ 
 is light while the other pair is  heavy ($\simeq M_{\rm GUT}$).  
The light Higgs doublets are identified as 
 the MSSM Higgs doublets ($H_u$ and $H_d$) and given by 
\begin{eqnarray} 
H_u &=& \widetilde{\alpha}_u  H_{10}^u 
+ \widetilde{\beta}_u  H_{126}^u \; , 
\nonumber \\
H_d &=& \widetilde{\alpha}_d  H_{10}^d  
+ \widetilde{\beta}_d  H_{126}^d  \; , 
\label{mix}
\end{eqnarray} 
where $\widetilde{\alpha}_{u,d}$ and $\widetilde{\beta}_{u,d}$ denote 
elements of the unitary matrix  which rotate the flavor basis in the 
original model into the (SUSY) mass eigenstates.  
Omitting the heavy Higgs mass eigenstates, the low energy 
superpotential is described by only the light Higgs doublets 
$H_u$ and $H_d$ such that 
\begin{eqnarray}
W_Y &=& 
(U^c) _i \left( \alpha^u  Y_{10}^{ij} + 
\beta^u Y_{126}^{ij} \right)  H_u \, Q_j 
+ (D^c)_i  
\left( \alpha^d  Y_{10}^{ij} + 
\beta^d Y_{126}^{ij}  \right) H_d \,Q_j  \nonumber \\ 
&+& (N^c)_i  
\left( \alpha^u  Y_{10}^{ij} -3 
\beta^u Y_{126}^{ij} \right)  H_u \,L_j 
+ (E^c)_i  
\left( \alpha^d  Y_{10}^{ij} -3 
\beta^d  Y_{126}^{ij}  \right) H_d \,L_j \nonumber \\ 
&+& 
  L_i \left( Y_{126}^{ij} \; v_T \right) L_j + 
 (N^c)_i  
  \left( Y_{126}^{ij} v_R \right)  (N^c)_j \; ,  
\label{Yukawa3}
\end{eqnarray} 
where the formulas of the inverse unitary transformation 
 of Eq.~(\ref{mix}), 
 $H_{10}^{u,d} = \alpha^{u,d} H_{u,d} + \cdots $ and 
 $H_{126}^{u,d} = \beta^{u,d} H_{u,d} + \cdots $, have been used. 

Providing the Higgs VEVs, $H_u = v \sin \beta$ and $H_d = v \cos \beta$ 
with $v = 174.1 \mbox{[GeV]}$, the quark and lepton mass matrices can be 
read off as
\begin{eqnarray}
 M_u &=& c_{10} M_{10} + c_{126} M_{126} \; , 
 \nonumber \\
 M_d &=& M_{10} + M_{126} \; ,   
 \nonumber \\
 M_D &=& c_{10} M_{10} - 3 c_{126} M_{126} \; , 
 \nonumber \\
 M_e &=& M_{10} - 3 M_{126} \; , 
 \nonumber \\
 M_T &=& c_T M_{126} \; ,
 \nonumber \\ 
 M_R &=& c_R M_{126} \;, 
\label{massmatrix}
\end{eqnarray} 
where $M_u$, $M_d$, $M_D$, $M_e$, $M_T$ and $M_R$ 
 denote the up-type quark, down-type quark, 
 neutrino Dirac, charged-lepton, 
 left-handed neutrino Majorana, 
 and right-handed neutrino Majorana mass matrices, respectively. 
Note that all the quark and lepton mass matrices 
 are characterized by only two basic mass matrices, $M_{10}$ and $M_{126}$,   
 and four complex coefficients 
 $c_{10}$, $c_{126}$, $c_T$ and $c_R$, 
 which are defined as 
 $M_{10}= Y_{10} \alpha^d v \cos\beta$, 
 $M_{126} = Y_{126} \beta^d v \cos\beta$, 
 $c_{10}= (\alpha^u/\alpha^d) \tan \beta$, 
 $c_{126}= (\beta^u/\beta^d) \tan \beta $, 
 $c_T = v_T/( \beta^d  v  \cos \beta)$) and 
 $c_R = v_R/( \beta^d  v  \cos \beta)$), respectively.  
These are the mass matrix relations required by 
 the minimal SO(10) model.

Low energy data of six quark masses, three mixing angles 
 and one phase in the Cabibbo-Kobayashi-Maskawa (CKM) matrix, 
 and three charged-lepton masses 
 are extrapolated to the GUT scale 
 according to the renormalization group equations (RGEs) 
 with given $\tan \beta$, 
 and the data set of quark and lepton mass matrices 
 at the GUT scale are obtained. 
Using the data, free parameters of fermion mass matrices 
 are determined so as to reproduce the neutrino oscillation data. 
As usually expected through the see-saw mechanism, 
 the mass scale of the right-handed neutrinos 
 is found to be around $M_R = 10^{12-13}$ GeV.

Note that in the minimal SO(10) model, 
 $Y_{126}$ is related to other fermion mass matrices 
 and determined so as to reproduce the fermion mass matrix data. 
Accordingly, the Higgs VEV, $v_R$, is determined 
 so as to provide the correct scale for the right-handed neutrino
 masses, which is found to be $v_R \simeq 10^{14}$ GeV 
 (see, for example, the second paper in Ref.~\cite{Goh:2003hf} 
  for the explicit presentations of the neutrino Dirac Yukawa matrix 
  and the right-handed neutrino mass matrix). 
This intermediate scale gives rise to the problem 
 on the gauge coupling unification discussed in Introduction.

In addition, as discussed by Chang {\it et al.} in Ref.~\cite{Chang:2004pb}, 
 when we introduce other Higgs multiples to break 
 the GUT gauge symmetry into the SM one, 
 beta function coefficients of RGEs of the gauge coupling 
 become very large and the gauge coupling quickly blows up 
 around the GUT scale. 
In field theoretical point of view, this implies that 
 the minimal SO(10) model should be defined as 
 an effective model with the cutoff around the GUT scale. 
The discrepancy between the GUT scale and the Planck scale 
 or the string scale, that would be a natural cutoff scale 
 of 4-dimensional field theory, can be understood 
 as a conceptual problem of the minimal SO(10) model.

\section{Minimal SO(10) model in a warped extra dimension} 
We consider a SUSY model 
 in the warped five dimensional brane world scenario \cite{RS}. 
The fifth dimension is compactified on the orbifold $S^1/Z_2$ 
 with two branes, ultraviolet (UV) and infrared (IR) branes, 
 sitting on each orbifold fixed point. 
With an appropriate tuning for cosmological constants 
 in the bulk and on the branes, 
 we obtain the warped metric \cite{RS}, 
\begin{eqnarray}
 d s^2 = e^{-2 k r_c |y|} \eta_{\mu \nu} d x^{\mu} d x^{\nu} 
 - r_c^2 d y^2 \; , 
\end{eqnarray}
 for $-\pi\leq y\leq\pi$, where $k$ is the AdS curvature, and 
 $r_c$ and $y$ are the radius and the angle of $S^1$, respectively.

By the compactification on the orbifold, 
 $N=1$ SUSY of the five dimensional theory, 
 which corresponds to $N=2$ SUSY in the four dimensional point of view, 
 is broken down to four dimensional $N=1$ SUSY. 
Supersymmetric Lagrangian of this system can be described 
 in terms of the superfield formalism 
 of four dimensional $N=1$ SUSY theories \cite{SUSYL1, SUSYL2, SUSYL3}. 
Now we consider the minimal SUSY SO(10) model in this warped geometry. 
There are lots of possibilities to construct such a model, 
 where some fields reside in the bulk and some reside 
 on the UV or the IR brane. 
The most important feature of the warped extra dimension model 
 is that the mass scale of the IR brane is warped down to 
 a low scale by the warp factor \cite{RS}, $ \omega = e^{-k r_c \pi}$, 
 in four dimensional effective theory. 
For simplicity, we take the cutoff of the original five dimensional theory 
 and the AdS curvature as 
 $M_5 \simeq k \simeq M_P=2.4 \times 10^{18}$ GeV, 
 the four dimensional (reduced) Planck mass,
 and so we obtain the effective cutoff scale 
 as $\Lambda_{IR}= \omega M_P$ in effective four dimensional theory. 
Now let us take the warp factor so as for the GUT scale 
 to be the effective cutoff scale 
 $ M_{\rm GUT}= \Lambda_{IR}=\omega M_P$, 
 namely $\omega \simeq 0.01$ \cite{Nomura:2006pn}. 
As a result, we can realize, as four dimensional effective theory, 
 the minimal SUSY SO(10) model 
 with the effective cutoff at the GUT scale.

Before going to a concrete setup of the minimal SO(10) model 
 in the warped extra dimension, 
 let us see Lagrangian for the hypermultiplet in the bulk, 
\begin{eqnarray}
{\cal L} &=& \int dy \left\{ 
\int d^4 \theta \; r_c \; e^{- 2 k r_c |y|} 
 \left( 
 H^{\dagger} e^{- V} H + H^{c} e^{ V}H^{c \dagger} 
 \right) \right. \nonumber \\
&+& 
\left. 
\int d^2 \theta e^{-3 k r_c |y|}
 H^{c} \left[ 
  \partial_{y} - \left( \frac{3}{2}-c \right) k r_c \epsilon(y) 
 - \frac{\chi}{\sqrt{2}}  \right]  
 H  +h.c. \right \} \; , 
\label{bulkL}
\end{eqnarray}
where $c$ is a dimensionless parameter, 
$\epsilon(y)=y/|y|$ is the step function, 
 $H, ~H^c$ is the hypermultiplet charged under some gauge group, 
 and 
\bea 
V &=& - \theta \sigma^\mu \bar{\theta} A_\mu
    -i \bar{\theta}^2 \theta \lambda_1  +i \theta^2 \bar{\theta} \bar{\lambda}_1
    + \frac{1}{2} \theta^2 \bar{\theta}^2 D \; , 
\nonumber \\
\chi &=&  \frac{1}{\sqrt{2}}(\Sigma + i A_5) +\sqrt{2} \theta\lambda_2
     + \theta^2 F \; , 
\eea
 are the vector multiplet and the adjoint chiral multiplets, 
 which form an $N=2$ SUSY gauge multiplet. 
 $Z_2$ parity for $H$ and $V$ is assigned as even, 
 while odd for $H^c$ and $\chi$.

When the gauge symmetry is broken down, 
 it is generally possible that the adjoint chiral multiplet 
 develops its VEV \cite{Kitano:2003cn}. 
Since its $Z_2$ parity is odd, 
 the VEV has to take the form, 
\bea
\left<\Sigma \right> = 2 \alpha k r_c  \epsilon(y) \; , 
\eea
where the VEV has been parameterized by a parameter $\alpha$. 
In this case, the zero mode wave function of $H$ 
 satisfies the following equation of motion:
\bea
\left[\partial_y -
 \left(\frac{3}{2}-c + \alpha \right) k r_c \epsilon(y) \right]H =0 
\eea
which yields 
\bea
H = \frac{1}{\sqrt{N}} 
 e^{ (3/2-c + \alpha) kr_c |y|} \; h(x^\mu) \; , 
\eea
where $h(x^\mu)$ is the chiral multiplet in four dimensions. 
Here, $N$ is a normalization constant 
 by which the kinetic term is canonically normalized, 
\be
\frac{1}{N} 
 =  \frac{(1-2 c+2 \alpha) k }
{e^{(1-2 c+2 \alpha) k r_c \pi}-1} \; . 
\ee
Hence, at $y=\pi$, the wave function becomes 
\bea
H(y=\pi) \simeq 
\sqrt{ (1-2 c + 2 \alpha ) k } \;  \omega^{-1} \; h(x^\mu) 
\eea
 if $e^{ (1/2 - c + \alpha ) k r_c \pi}  \gg 1$, while 
\bea 
H(y=\pi) \simeq 
 \sqrt{- (1-2 c + 2 \alpha ) k } \;  \omega^{-1} 
 e^{ (1/2 - c +  \alpha ) k r_c \pi}  \; h(x^\mu) 
\eea 
 for $ e^{ (1/2 -  c +  \alpha ) k r_c \pi}  \ll 1$. 

Lagrangian for a chiral multiplets on the IR brane is given by 
\bea 
 {\cal L}_{IR}=  
 \int d^4 \theta \;  \omega^\dagger \omega \;  \Phi^\dagger \Phi 
 +\left[   \int d^2 \theta \;  \omega^3 \; W(\Phi) + h.c.   
 \right] \; ,
\eea 
where we have omitted the gauge interaction part 
 for simplicity. 
If it is allowed by the gauge invariance, 
 we can write the interaction term 
 between fields in the bulk and on the IR brane, 
\bea 
{\cal L}_{int}= \int d^2 \theta \omega^3 
 \frac{Y}{\sqrt{M_5}} \Phi^2 H(y=\pi) +h.c. \; ,  
\label{IR-Yukawa} 
\eea  
where $Y$ is a Yukawa coupling constant, 
 and $M_5$ is the five dimensional Planck mass 
 (we take $M_5\sim M_P$ as mentioned above, for simplicity). 
Rescaling the brane field $\Phi \rightarrow  \Phi/\omega$ 
 to get the canonically normalized kinetic term 
 and substituting the zero-mode wave function of the bulk fields, 
 we obtain Yukawa coupling constant 
 in effective four dimensional theory as 
\bea 
  Y_{4D} \sim Y 
\eea 
 if $e^{ (1/2 - c +  \alpha ) k r_c \pi}  \gg 1$,  
 while 
\bea 
  Y_{4D} \sim Y 
  \times e^{ (1/2 -  c +  \alpha ) k r_c \pi}  \ll Y \; , 
\label{suppression}
\eea  
 for $e^{ (1/2 - c + \alpha ) k r_c \pi }  \ll 1$. 
In the latter case, we obtain a suppression factor 
 since $H$ is localized around the UV brane.

Now we give a simple setup of the minimal SO(10) model 
 in the warped extra dimension. 
We put all ${\bf 16}$ matter multiplets on the IR ($y=\pi$) brane, 
 while the Higgs multiplets ${\bf 10}$ and $\overline{\bf 126}$ 
 are assumed to live in the bulk. 
In Eq.~(\ref{IR-Yukawa}), replacing the brane field into the matter 
 multiplets and the bulk field into the Higgs multiplets, 
 we obtain Yukawa couplings in the minimal SO(10) model. 
The Lagrangian for the bulk Higgs multiplets are given 
 in the same form as Eq.~(\ref{bulkL}), 
 where $\chi$ is the SO(10) adjoint chiral multiplet, ${\bf 45}$. 
As discussed above, since the SO(10) gauge group is broken 
 down to the SM one, 
 some components in $\chi$ which is singlet under the SM gauge group 
 can in general develop VEVs. 
Here we consider a possibility that 
 the ${\rm U}(1)_X$ component 
 in the adjoint $\chi ={\bf 45}$ under the decomposition 
 SO(10) $\supset {\rm SU}(5) \times {\rm U}(1)_X$ has 
 a non-zero VEV\footnote{
Since $\chi$ has an odd $Z_2$ parity, its non-zero VEV leads to 
 the Fayet-Iliopoulos D-terms localized 
 on both the UV and IR branes \cite{D-term}, 
 which should be canceled to preserve SUSY. 
For this purpose, we have to introduce new fields on both branes 
 by which the D-terms are compensated. 
If such fields are in the same representations 
 as matters or Higgs fields like ${\bf 16}$ or $\overline{\bf 126}$, 
 we would need to impose some global symmetry to distinguish them. 
}, 
\bea
 {\bf 45} = {\bf 1}_0
 \oplus {\bf 10}_{+4} \oplus \overline{\bf 10}_{-4}
 \oplus {\bf 24}_0 \; .   \nonumber 
\eea
The ${\bf 10}$ Higgs multiplet and 
 the $\overline{\bf 126}$ Higgs multiplet 
 are decomposed under ${\rm SU}(5) \times {\rm U}(1)_X$ as
\bea
{\bf 10} &=& {\bf 5}_{+2} \oplus \overline{\bf 5}_{-2} \; , 
 \nonumber \\
\overline{\bf 126} &=& 
 {\bf 1}_{+10} 
 \oplus {\bf 5}_{+2} \oplus \overline{\bf 10}_{+6}
 \oplus {\bf 15}_{-6} 
 \oplus \overline{\bf 45}_{-2} \oplus {\bf 50}_{+2} \; . 
 \nonumber  
\eea
In this decomposition, 
 the coupling between a bulk Higgs multiplet and 
 the ${\rm U}(1)_X$ component in $\chi$ is proportional 
 to  ${\rm U}(1)_X$ charge, 
\be
{\cal L}_{int} \supset \frac{1}{2} \int d^2 \theta \omega^3
Q_X \langle \Sigma_X \rangle H^c H + h.c. \;,
\ee
 and thus each component effectively obtains 
 the different bulk mass term,  
\bea 
  \left( \frac{3}{2} - c \right) k r_c  
   + \frac{1}{2}Q_X \langle \Sigma_X \rangle,   
\eea
 where $Q_X$ is the ${\rm U}(1)_X$ charge of corresponding Higgs multiplet, 
 and $\Sigma_X$ is the scalar component of 
 the ${\rm U}(1)_X$ gauge multiplet (${\bf 1}_0$). 
Now we obtain different configurations of the wave functions 
 for these Higgs multiplets. 
Since the ${\bf 1}_{+10}$ Higgs has a large ${\rm U}(1)_X$ charge 
 relative to other Higgs multiplets, 
 we can choose parameters $c$ and $\langle \Sigma_X \rangle$ 
 so that Higgs doublets are mostly localized around the IR brane 
 while the ${\bf 1}_{+10}$ Higgs is localized around the UV brane. 
For example, the parameter choice, 
 $c=-7/2$ for both ${\bf 10}$ and $\overline{\bf 126}$ Higgs multiplets 
 and $ \langle \Sigma_X \rangle = - k r_c$, 
 can realize this situation.

Using the decomposition of matter multiplets, 
\bea
 {\bf 16}^i = {\bf 1}_{-5}^i 
 \oplus \overline{\bf 5}_{+3}^i  \oplus {\bf 10}_{-1}^i  \; ,
 \nonumber  
\eea
the Yukawa couplings between matters and 
 the $\overline{\bf 126}$ Higgs multiplet 
 on the IR brane are decomposed into  
\bea
 W_{Y_{126}} &=& 
  Y_u^{ij} {\bf 5}_{+2} {\bf 10}_{-1}^i {\bf 10}_{-1}^j 
+ Y_d^{ij} \overline{\bf 45}_{-2} 
  \overline{\bf 5}_{+3}^i {\bf 10}_{-1}^j \nonumber \\
&+& Y_D^{ij} {\bf 5}_{+2} {\bf 1}_{-5}^i \overline{\bf 5}_{+3}^j 
 + Y_e^{ij} \overline{\bf 45}_{-2} \overline{\bf 5}_{+3}^i {\bf 10}_{-1}^j
\nonumber \\ 
&+& 
  Y_{\nu_L}^{ij} {\bf 15}_{-6}
   \overline{\bf 5}_{+3}^i  \overline{\bf 5}_{+3}^j 
+ Y_{\nu_R}^{ij} {\bf 1}_{+10} {\bf 1}_{-5}^i {\bf 1}_{-5}^j \; . 
\eea
Here, all the Yukawa couplings coincide with 
 the original Yukawa coupling $Y_{126}$ 
 up to appropriate CG coefficients. 
As discussed above, the ${\bf 1}_{+10}$ Higgs multiplet 
 giving masses for right-handed neutrinos
 is localized around the UV brane and, therefore, 
 we obtain a suppression factor 
 as in Eq.~(\ref{suppression}) 
 for the effective Yukawa coupling between 
 the Higgs and right-handed neutrinos. 
In effective four dimensional description, 
 the GUT mass matrix relation is partly broken down, 
 and the last term in Eq.~(\ref{Yukawa3}) is replaced into 
\bea 
 Y_{126}^{ij} v_R \rightarrow Y_{126}^{ij} (\epsilon v_R) \; ,  
\eea
where $\epsilon$ denotes the suppression factor. 
By choosing appropriate parameters 
 so as to give $\epsilon \simeq 10^{-2}$, 
 we can take $v_R \simeq M_{\rm GUT}$ 
 and keep the successful gauge coupling unification in the MSSM. 
In fact, the above parameter set, 
 $c=-7/2$ and $ \langle \Sigma_X \rangle = - k r_c$, 
 leads to $\epsilon = \omega = M_{\rm GUT}/M_P \simeq 10^{-2}$. 
The other Higgs multiplets are localized around the IR brane, 
 so that there is no suppression factor for other effective 
 Yukawa couplings.

In our setup, all the matters reside on the brane 
 while the Higgs multiplets reside in the bulk. 
This setup shares the same advantage as 
 the so-called orbifold GUT \cite{Kawamura:2000ev,
Altarelli:2001qj, Hall:2001pg}. 
We can assign even $Z_2$ parity 
 for MSSM doublet Higgs superfields 
 while odd for triplet Higgs superfields, 
 as a result, the proton decay process through 
 dimension five operators are forbidden.

\section{Conclusion}
The minimal renormalizable supersymmetric SO(10) model 
 is a simple framework to reproduce current data 
 for fermion masses and flavor mixings with some predictions. 
However, this model suffers from some problems related to 
 the running of the gauge couplings. 
To fit the neutrino oscillation data, 
 the mass scale of right-handed neutrinos lies 
 at the intermediate scale. 
This implies the presence of some Higgs multiplets 
 lighter than the GUT scale. 
As a result, the gauge coupling unification in the MSSM 
 may be spoiled. 
In addition, since Higgs multiplets of large representations 
 are introduced in the model, 
 the gauge couplings blow up around the GUT scale. 
Thus, the minimal SO(10) model would be effective theory 
 with a cutoff around the GUT scale, far below the Planck scale. 

In order to solve these problems, we have considered 
 the minimal SO(10) model in the warped extra dimension. 
As a simple setup, we have assumed that matter multiplets 
 reside on the IR brane 
 while the Higgs multiplets reside in the bulk. 
The warped geometry leads to a low scale effective cutoff
 in effective four dimensional theory, 
 and we fix it at the GUT scale. 
Therefore, the four dimensional minimal SO(10) model 
 is realized as the effective theory with the GUT scale cutoff.

After the GUT symmetry breaking, the adjoint scalar in  
 the gauge multiplet in five dimensional SUSY 
 can generally develop a VEV, 
 which plays a role of bulk mass for the bulk Higgs multiplets. 
This bulk mass is proportional to the charge 
 of each Higgs multiplets and cause the difference 
 between wave functions of each Higgs multiplet. 
We have found the possibility that 
 the singlet Higgs which provides right-handed neutrino with masses 
 is localized around the UV brane and the geometrical suppression factor 
 emerges in Yukawa couplings of the right-handed neutrinos. 
As a result, we can set the mass scale of the right-handed neutrinos 
 at the intermediate scale 
 nevertheless the singlet Higgs VEV is around the GUT scale. 
All Higgs multiplets naturally have masses around the GUT scale 
 and the gauge coupling unification in the MSSM remains the same.

Finally, we give some comments. 
One can easily extend our setup to put some 
 of matter multiplets in the bulk 
 \cite{Grossman:1999ra, Chang:1999nh, Gherghetta:2000qt}. 
In this case, we may explain the fermion mass hierarchy 
 in terms of the different overlapping of 
 fermion wave functions between different generations. 
In this paper, we have assumed that the GUT gauge symmetry 
 is successfully broken down to the SM one. 
There are several possibilities for the GUT symmetry breaking. 
It is easy to introduce appropriate Higgs multiplets 
 and superpotential so as to break the GUT symmetry 
 on a brane as in four dimensional SO(10) models. 
We also can introduce an appropriate boundary  conditions 
 for bulk gauge multiplets to (explicitly) break 
 the GUT symmetry to a subgroup with rank five in total, 
 as in the orbifold GUTs.

\section*{Acknowledgments}
We would like to thank Siew-Phang Ng for useful comments 
 and reading the manuscript. 
The work of T.F. and N.O. is supported in part by 
 the Grant-in-Aid for Scientific Research from the Ministry 
 of Education, Science and Culture of Japan (\#16540269, \#18740170).


\end{document}